\documentclass[aps,twocolumn,prl,superscriptaddress]{revtex4}

\usepackage{mathptmx}
\usepackage{graphicx}

\begin{document}

\title{Rogue waves in the atmosphere}

\author{Lennart Stenflo}
\affiliation{Department of Physics, Link\"oping University, SE--581 83 Link\"oping, Sweden} 

\author{Mattias Marklund}
\email{mattias.marklund@physics.umu.se}
\affiliation{Department of Physics, Ume{\aa} University, SE--901 87 Ume{\aa}, Sweden}

\begin{abstract}
	The appearance of rogue waves is well known in oceanographics, optics, and cold matter systems. Here we show a possibility for the existence of atmospheric rogue waves.
\end{abstract}

\maketitle

Solitary waves in plasmas, and similarly in the
atmosphere, have been
thoroughly investigated during more than four
decades \cite{1}. Observations
\cite{2,3} of atmospheric disturbances propagating
with no appreciable change
in structure are supported by theoretical
descriptions where the vector
nonlinearities in the atmospheric wave equations
play a dominant role.
Acoustic-gravity vortices in the atmosphere have
thus been found \cite{4}.

Related studies of waves in the oceans have
recently focused on the
appearance of oceanic rogue waves (e.g. \cite{5,6}).
Nonlinear wave studies related to the rogue wave phenomena can also be found in optics \cite{optics}, superfluid-He4 \cite{superfluid}, and optical cavities \cite{cavity}.
Whereas theoretical models
can now fairly well describe oceanic rogue waves
\cite{7} there is, as far as we
know, no corresponding theory for the
atmospheric rogue waves that have
been observed \cite{8}. We shall therefore here point
out a way to describe
such atmospheric waves.

One should then start with the basic theory for
atmospheric waves that is
described in, for example, a recent review paper
on nonlinear
acoustic-gravity waves \cite{9}. In contrast to most
previous two-dimensional
studies where the vector nonlinearities in the
equations dominate, and
where modon solutions thus can occur, we shall
here look at the
one-dimensional propagation along one of the
horizontal directions (the
$x$-direction). In that case only the scalar
nonlinearity remains, and the
slowly varying amplitude $A(x,t)$ of the
normalized pressure perturbation is
consequently governed by the nonlinear
Schr\"odinger equation \cite{10}
\begin{equation}
	2iK\frac{\partial A}{\partial x} + f\frac{\partial^2A}{\partial\tau^2} 
	- \frac{2K^2n_1|A|^2}{n_0}A = 0		
\end{equation}
where $\tau = t - x/v_0$, $v_0$ is the linear group velocity, 
$K = \omega n_0/c$ is the wavenumber, $f = ({K}/{v_0^2}){\partial v_0}/{\partial\omega}$ is the group velocity dispersion,  
and $n_0$ is the linear refractive index (here the nonlinear refractive index is given by $n_0 + n_1|A|^2$, see Ref.\ \cite{10}). We next
normalize our variables, so that $x \rightarrow -Kx$, $t \rightarrow K\tau/|f|^{1/2}$, and $A \rightarrow (n_1/n_0)^{1/2}A$, to obtain
\begin{equation}
	i \frac{\partial A}{\partial x} + \frac{1}{2}\frac{\partial^2 A}{\partial t^2} + |A|^2A = 0 
\end{equation}
which, for example, has the rogue wave solution (see Fig. 1)
\begin{equation}
	A(x,t) = \left( 4\frac{1 + 2ix}{1 + 4x^2 + 4t^2} - 1\right)\exp(ix)
\end{equation}
\begin{figure}
	\includegraphics[width=0.8\columnwidth]{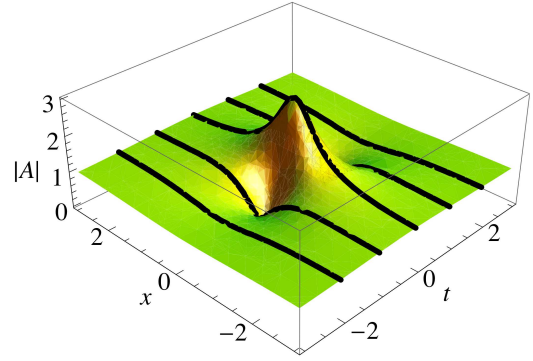}
	\caption{The evolution of the absolute value of the solution (3).}
\end{figure}

It is now clearly established that the ocean rogue wave phenomenon is much more common than expected from early wave modeling \cite{11}. The description of ocean rogue waves by the nonlinear Schr\"odinger equation \cite{12} bears close resemblance to the dito for atmospheric disturbances. Encounters of rogue waves on the ocean is a truly horrific experience, and it would of course be an even greater threat to meet the atmospheric counterpart while flying. Investigations of these nonlinear wave structures in the atmosphere could also be of interest for a wide community of researchers, from both basic and applied sciences.




\end{document}